\documentclass[aps, showpacs,pra,twocolumn] {revtex4}
\usepackage{hyperref}
\usepackage{amsmath }
\usepackage{amssymb}
\usepackage{epsfig}
\usepackage{natbib}
\newcommand*{\be}{\begin{equation}}
\newcommand*{\ee}{\end{equation}}

\begin{document}
\bibliographystyle{revtex}
\title{Exact wave functions of electron in a quantum dot \\ with account of  the Rashba
spin-orbit interaction}
\author{V. V. Kudryashov}
\email{kudryash@dragon.bas-net.by}

\affiliation{Institute of Physics, National Academy of Sciences of
Belarus, \\ 68 Nezavisimosti  Ave., 220072, Minsk,  Belarus}

\begin{abstract}
 We present the exact wave functions and energy levels
of electron in a two-dimensional circular quantum dot in the presence of the Rashba
spin-orbit interaction. The confinement is described by the realistic potential well of finite depth.
\end{abstract}


\maketitle

The Rashba spin-orbit interaction in semiconductor quantum dots has been the object
of many investigations in recent years (see \cite{bul, tsi, val, kua} and references therein).
The Rashba interaction is of the form \cite{ras,byc}
\begin{equation}
V_R= \beta_R (\sigma_x p_y - \sigma_y p_x)
\end{equation}
with standard Pauli spin-matrices
\begin{equation}
  \sigma_x = \left(\begin{array}{cc}
 0&1 \\
 1&0
 \end{array}\right), \quad
\sigma_y =  \left(\begin{array}{cr}
 0&-i \\
 i&0
 \end{array}\right) .
\end{equation}
The Rashba interaction can be strong in semiconductor heterostructures  and its strength can be
controlled by an external electric field.

The quantum dots in semiconductors can be described as effectively
two-dimensional systems in confining potential $V_c(x,y)$.
 A confining potential is usually assumed to be
symmetric, $V_c(x,y)= V_c(\rho), \rho =\sqrt{x^2 +y^2}$. Then the
single-electron wave functions satisfy  the Schr\"odinger equation
\begin{equation}
 \left(\frac{{\bf p}^2}{2 \mu} + V_{c}(\rho) +V_R\right) \Psi = E \Psi ,
\end{equation}
where $\mu$ is the effective electron mass.  There are two model
potentials which are widely employed in this area. The first is a
harmonic oscillator potential. Such a model  \cite{val,kua} admits
the approximate (not exact) solutions of Eq.~(3).  The second model \cite{bul,tsi} is a
circular quantum dot with hard walls
 $ V_{c}(\rho)=0$ for $\rho <  \rho_0$, $
V_{c}(\rho)= \infty$ for $\rho >  \rho_0$.
 This model is  exactly solvable. In the framework of above models  the number
of  allowed energy levels is infinite for the fixed total angular momentum.

In this paper, we examine a model  which corresponds to a circular quantum dot with a rectangular
potential well of finite depth: $V_c(\rho) =0$ for $\rho <  \rho_0$, $
V_{c}(\rho)= V_0 >~0$ for $\rho >  \rho_0$ . We shall present the exact coordinate wave functions and
energy levels for the wells of arbitrary depth.

The analogous model with replacement $V_c(\rho) \to V_c(\rho) - V_0$ was proposed in \cite{chap},
where  the approximate investigation was developed in the momentum representation for a shallow well.
As to the coordinate representation, unfortunately  the paper \cite{chap} does not contain the radial Schr\"odinger
equations as well as  the explicit and complete expressions for solutions of these equations. Emphasizing
different aspects of the same problem the present paper and the paper \cite{chap} mutually complement each other.

 The Schr\"odinger equation
 is considered  in the cylindrical coordinates
 $x = \rho \cos \varphi,  y = \rho \sin \varphi$.
 Further it
is convenient to employ dimensionless  quantities
 \begin{equation}
 e =\frac{2 \mu}{\hbar^2} \rho_0^2 E, \quad
  v= \frac{2 \mu}{\hbar^2} \rho_0^2 V_0, \quad
 \beta=\frac{2 \mu}{\hbar} \rho_0 \beta_R ,  \quad r= \frac{\rho}{
 \rho_0}.
 \end{equation}
As it was shown in \cite{bul}, Eq. (3) permits the separation of
variables:
\begin{equation}
 \Psi_m(r,\varphi) =  u(r) e^{i m \varphi}\left(\begin{array}{c}
 1 \\
  0
 \end{array}\right) + w(r)  e^{i (m+1) \varphi} \left(\begin{array}{c}
 0 \\
  1
 \end{array}\right) ,
\end{equation}
\[
m=0, \pm 1, \pm 2, \ldots
\]
due to conservation of the total angular momentum
$L_z + \frac{\hbar}{2} \sigma_z $.

In \cite{bul,tsi}, the requirements $ u(1)= w(1)=0 $
 were imposed. In the present model, we look for the radial wave functions
$u(r)$ and $w(r)$ regular at  the origin $r=0$ and decreasing at
infinity $r \rightarrow \infty$.

We consider two regions $ r <1$ (region 1) and $r >1$  (region 2) separately.

In the region 1 $(v=0)$ the radial equations may be written in the suitable form
\begin{eqnarray}
r^2 \frac{d^2u_1}{dr^2} & + &r \frac{d u_1}{d r}  +( k^+_1 k^-_1 r^2 - m^2)u_1
\nonumber \\
&= &( k^+_1 -  k^-_1) r^2\left(\frac{d w_1}{d r} +  \frac{m+1}{r} w_1 \right) ,
\nonumber \\
 r^2\frac{d^2w_1}{dr^2} &+ &r \frac{d w_1}{d r}  +( k^+_1 k^-_1 r^2 -( m+1)^2)w_1
 \nonumber \\
&= &- ( k^+_1 -  k^-_1) r^2 \left(\frac{d u_1}{d r}- \frac{m}{r} u_1 \right) ,
\end{eqnarray}
where
\begin{equation}
k^{\pm}_1(e,\beta) = \sqrt{e +\frac{\beta^2}{4}} \pm
\frac{\beta}{2} .
\end{equation}
Note that in the case $\beta = 0$ ($k^+_1  = k^-_1$)  Eqs. (6) are the Bessel equations.
Therefore, following  \cite{bul,tsi} we use the known properties \cite{abr}
\begin{eqnarray}
  \left(\frac{d}{d r} - \frac{n}{r} \right) J_n(k r)
 &= &-k J_{n+1}(k r)  , \nonumber \\
  \left(\frac{d}{d r} + \frac{n}{r} \right) J_n(k r)
 &= &k J_{n-1}(k r)
 \end{eqnarray}
 of the Bessel functions  in order to obtain the exact solutions
\begin{eqnarray}
u_1(m,e,\beta,r) & =& c_1 f_1(m,e,\beta,r) \nonumber \\ &+& d_1 g_1(m,e,\beta,r),
\nonumber \\
 w_1(m,e,\beta,r)  &= &c_1g_1(m+1,e,\beta,r)  \nonumber \\ &+& d_1 f_1(m+1,e,\beta,r)
\end{eqnarray}
 of system (6), where
 \begin{eqnarray}
 f_1(m,e,\beta,r) &= &\frac{1}{2}\left(J_m(k^-_1r) + J_m(k^+_1r) \right),
 \nonumber \\
g_1(m,e,\beta,r)& = &\frac{1}{2}\left(J_m(k^-_1r) - J_m(k^+_1r)
\right)
\end{eqnarray}
are the real linear combinations of the Bessel functions with real
arguments.
Here $c_1$ and $d_1$ are arbitrary coefficients.
The radial wave functions $u_1(r)$ and $w_1(r)$ have the desirable behavior at
the origin.

 In the region 2 $(v>0)$ we have the  radial equations
\begin{eqnarray}
r^2 \frac{d^2u_2}{dr^2} &+& r \frac{d u_2}{d r}  -(k^+_2 k^-_2  r^2 +
m^2)u_2  \nonumber \\
&= &- i ( k^+_2 -  k^-_2) r^2\left(\frac{d w_2}{d r} +  \frac{m+1}{r} w_2
\right), \nonumber \\
 r^2\frac{d^2w_2}{dr^2} &+& r \frac{d w_2}{d r}  -(k^+_2 k^-_2 r^2 +(
 m+1)^2)w_2
 \nonumber \\
&= &i( k^+_2 -  k^-_2)  r^2 \left(\frac{d u_2}{d r}- \frac{m}{r} u_2 \right) ,
\end{eqnarray}
 where
\begin{equation}
k^{\pm}_2(e,v,\beta) = \sqrt{v-e -\frac{\beta^2}{4}} \pm i
\frac{\beta}{2}.
\end{equation}
Now in the case $\beta = 0$ ($k^+_2  = k^-_2$)  Eqs. (11) are the modified Bessel equations.
Therefore, using the known properties \cite{abr}
\begin{eqnarray}
 \left(\frac{d}{d r} - \frac{n}{r} \right) K_n(k r)
 &= &-k K_{n+1}(k r) , \nonumber \\
  \left(\frac{d}{d r} + \frac{n}{r} \right) K_n(k r)
 &=& -k K_{n-1}(k r)
\end{eqnarray}
of the modified Bessel functions
 it is easily to get the exact solutions
\begin{eqnarray}
u_2(m,e,v,\beta,r) & =&
 c_2 f_2(m,e,v,\beta,r)  \nonumber \\&+& d_2
g_2(m,e,v,\beta,r), \nonumber \\
 w_2(m,e,v,\beta,r)  & =&
 c_2 g_2(m+1,e,v,\beta,r) \nonumber \\ &- &d_2 f_2(m+1,e,v,\beta,r)
\end{eqnarray}
 of system (11), where
 \begin{eqnarray}
 f_2(m,e,v,\beta,r)& =& \frac{1}{2}\left(K_m(k^-_2r) + K_m(k^+_2r) \right),\nonumber \\
 g_2(m,e,v,\beta,r)& = &\frac{i}{2}\left(K_m(k^-_2r) - K_m(k^+_2r)
\right)
\end{eqnarray}
are the real linear combinations of the modified Bessel functions
with complex arguments.
Here $c_2$ and $d_2$ are arbitrary coefficients.
At large values of $r$ the functions $ f_2(m,e,v,\beta,r)$ and $ g_2(m,e,v,\beta,r)$ behave as
\begin{eqnarray}
f_2(m,e,v,\beta,r) &\sim & \sqrt{\frac{\pi}{2}}\frac{1}{(v-e)^{1/4}} \nonumber \\
&\times&  \frac{e^{-r \sqrt{v-e -\beta^2/4} }}{\sqrt{r}}
\cos\left(\frac{\beta r +\gamma}{2} \right) , \nonumber \\
g_2(m,e,v,\beta,r) &\sim &  -\sqrt{\frac{\pi}{2}}\frac{1}{(v-e)^{1/4}} \nonumber \\
&\times&   \frac{e^{-r \sqrt{v-e -\beta^2/4} }}{\sqrt{r}}
\sin\left(\frac{\beta r +\gamma}{2}\right) ,
\end{eqnarray}
where $\gamma$ is defined as follows,
\begin{equation}
\cos(\gamma)= \frac{\sqrt{v-e -\beta^2/4}}{\sqrt{v-e }} , \quad
\sin(\gamma) = \frac{\beta}{2 \sqrt{v-e }} .
\end{equation}
Thus, the radial wave functions $u_2(r)$ and $w_2(r)$ have the appropriate behavior at
infinity. It should be noted that the radial wave functions have  the infinite number of zeros
 for the finite value of energy.

 Since the solutions (9) are valid for $e > -\beta^2/4$ and the solutions (14) are valid for $e < v -\beta^2/4$  then
 the complete energy range is $-\beta^2/4 < e < v -\beta^2/4$.

The continuity conditions
 \begin{eqnarray}
u_1(m,e,\beta,1)& = &u_2(m,e,v,\beta,1), \nonumber \\
  u'_1(m,e,\beta,1)&=& u'_2(m,e,v,\beta,1), \nonumber \\
 w_1(m,e,\beta,1) &= &w_2(m,e,v,\beta,1), \nonumber \\
  w'_1(m,e,\beta,1)&=& w'_2(m,e,v,\beta,1)
\end{eqnarray}
for the radial wave functions and their derivatives at the boundary point $r=1$ lead to the algebraic
equations
\begin{equation}
M(m,e,v,\beta) \left(\begin{array}{c}
 c_1 \\
 c_2 \\
 d_1 \\
 d_2
   \end{array}\right) =0
\end{equation}
for coefficients $c_1, c_2, d_1$ and  $d_2$, where

\begin{table*}[t]
\caption{ Energy levels.}\label{t1}
%\begin{center}
\begin{tabular}{l r r r r r r r r} \hline \hline
$v$&$\beta$& &  &e  & &&
\\ \hline
 \multicolumn{9}{c}{$m=0$}  \\ \hline
 $25$&$0$&$3.98$&$9.94$&$19.61$&  &  &  & \\
 $25$&$1$&$3.49$&$9.81$&$19.15$&  &  &  & \\
 $25$&$5$&$-4.40$&$2.83$&$13.40$&  &  & & \\
 $25$&$10$&$-23.25$&$-18.31$&$-9.67$ & & &
 \\ \hline
 $49$&$0$&$4.41$&$11.13$&$22.75$&$35.91$&  &  & \\
 $49$&$1.4$&$3.49$&$11.03$&$21.87$&$35.79$&  &  & \\
 $49$&$7$&$-10.40$&$-3.71$&$9.55$&$24.22$&  & \\
 $49$&$14$&$-47.11$&$-41.45$&$-32.19$&$-19.60$&$-2.25$ &
 \\ \hline
 $100$&$0$ &$4.77$  &$12.09$&$24.97$&$40.08$&$60.28$&$81.84$ &\\
 $100$&$2$ &$2.97$  &$11.75$&$23.30$&$39.73$&$58.65$&$81.42$ &\\
 $100$&$10$&$-21.91$&$-16.95$&$-4.88$&$14.82$&$35.04$&$56.69$ & \\
 $100$&$20$&$-97.96$&$-91.83$&$-81.75$&$-67.58$&$-49.91$&$-28.53$&$-1.66$
 \\ \hline
  \multicolumn{9}{c}{$m=1$}  \\ \hline
 $25$&$0$&$9.94$&$17.46$&  &  &  &  &\\
 $25$&$1$&$9.02$&$17.85$&  &  &  &  &\\
 $25$&$5$&$-2.52$&$9.26$&  &  &  &  &\\
 $25$&$10$&$-23.22$&$-17.29$&$-4.85$  &  &  &
 \\ \hline
 $49$&$0$&$11.13$&$19.85$&$35.91$&  &  &  &\\
 $49$&$1.4$&$9.41$&$20.50$&$34.28$&  &  &  &\\
 $49$&$7$&$-9.62$&$1.84$&$20.95$&$36.73$ & & \\
 $49$&$14$&$-47.04$&$-41.12$&$-31.51$&$-13.33$   & &
 \\ \hline
 $100$&$0$&$12.09$&$21.66$&$40.08$&$57.25$&$81.84$&  & \\
 $100$&$2$&$ 8.85$&$22.59$&$37.08$&$58.13$&$79.02$&  & \\
 $100$&$10$&$-22.91$&$-14.79$&$4.24$&$31.61$&$55.83$&$74.93$ & \\
 $100$&$20$&$-97.91$&$-91.72$&$-81.26$&$-66.83$&$-48.24$&$-17.86$ &
 \\ \hline
  \multicolumn{9}{c}{$m=2$}  \\ \hline
 $25$&$0$&$17.46$&  &  &  &  &  &\\
 $25$&$1$&$16.13$&  &  &  &  &  &\\
 $25$&$5$&$1.30$& $16.08$ &  &  &  & & \\
 $25$&$10$&$-22.86$&$-13.35$&$-0.20$& & &
 \\ \hline
 $49$&$0$&$19.85$&$30.35$&  &  &  &  & \\
 $49$&$1.4$&$17.37$&$31.72$&  &  &  &  & \\
 $49$&$7$&$-6.88$&$10.11$&$32.20$  &  &  & & \\
 $49$&$14$&$-44.83$&$-40.77$&$-26.05$&$-4.15$ & &
 \\ \hline
 $100$&$0$&$21.66$&$33.34$&$57.25$&$76.20$&  &   &\\
 $100$&$2$&$17.08$&$35.51$&$52.95$&$78.21$&  & & \\
 $100$&$10$&$-22.12$&$-8.70$&$16.99$&$49.86$&$74.91$ & &\\
 $100$&$20$&$-97.84$&$-91.39$&$-80.29$&$-65.52$&$-37.69$&$-3.45$ &
  \\ \hline \hline
\end{tabular}
%\end{center}
\end{table*}

\begin{widetext}
 \begin{equation}
 M(m,e,v.\beta)=
  \left(\begin{array}{rrrr}
  f_1(m,e,\beta,1)&-f_2(m,e,v,\beta,1)&
   g_1(m,e,\beta,1)&-g_2(m,e,v,\beta,1) \\
  f'_1(m,e,\beta,1)&-f'_2(m,e,v,\beta,1)&
   g'_1(m,e,\beta,1)&-g'_2(m,e,v,\beta,1) \\
 g_1(m+1,e,\beta,1)&-g_2(m+1,e,v,\beta,1)&
   f_1(m+1,e,\beta,1)&f_2(m+1,e,v,\beta,1) \\
 g'_1(m+1,e,\beta,1)&-g'_2(m+1,e,v,\beta,1)&
   f'_1(m+1,e,\beta,1)&f'_2(m+1,e,v,\beta,1)
 \end{array}\right) .
\end{equation}
\end{widetext}

Hence, the exact equation for energy spectrum is
\begin{equation}
\det M(m,e,v,\beta) =0.
\end{equation}
 It should be stressed that in the explored model
the number of  admissible energy levels is finite for the fixed total angular momentum.
Note that Eq. (21) is invariant under two replacements $ m \to
-(m+1)$ and $\beta \rightarrow -\beta$.

 \begin{figure}[t]% Fig.1.
\centering
\includegraphics{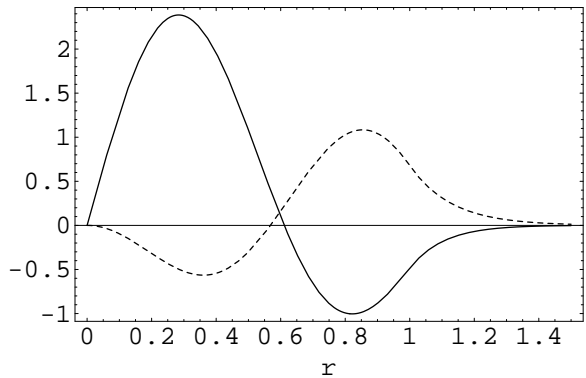} \caption{Radial wave functions  for $m=1$. Solid line for $u(r)$, dashed line for $w(r)$. }\label{f1}
\end{figure}
\begin{figure}[t]% Fig.1.
 \centering
\includegraphics{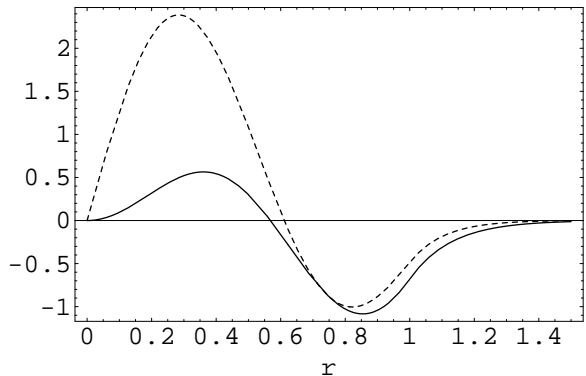} \caption{Radial wave functions   for $m= -2$. Solid line for $u(r)$, dashed line for $w(r)$.}\label{f2}
\end{figure}

 If the energy values are found from Eq. (21) then it is simply to get
 the values of coefficients   $c_1, c_2, d_1$ and $d_2$ from Eq. (19) and the following normalization
 condition
 $
 \int_0^{\infty}\left(u^2(r) +  w^2(r)\right)r dr = 1 .
 $

Now we present some numerical and graphic illustrations in
addition to the analytic results.

Table \ref{t1} shows the dependence of energy $e$ on the Rashba parameter $\beta$,  the well depth $v$
 and angular momentum
number $m$. The values of $\beta$ are $0, 0.2 \sqrt{v}, \sqrt{v}$ and $2 \sqrt{v}$ for the given value of $v$. We see that the number of energy levels decreases  if
the well depth $v$ decreases and if angular momentum number $m$
increases. Besides we see that the negative energy values appear for the positive $v$ when the values of $\beta$ becomes sufficiently large  in comparison with the value of $\sqrt{v}$. If  $\beta^2/4 \geq v$ then the
energy spectrum is completely shifted to negative region.

Figures \ref{f1} and \ref{f2} demonstrate the  continuous radial wave functions for $m=1$ and $m=-2$  respectively
while $\beta=2, v=100$ and $e=37.0825$ in both cases. The solid lines correspond to
the functions $u(r)$ and the dashed lines correspond to the functions $w(r)$. The numerical values of coefficients
are $ c_1 =4.22035, c_2 =-4067.87, d_1 =-0.7139284$ and $d_2 = 880.843$ in the case of Fig. 1 and
  $ c_1 = 0.713928, c_2 = 880.843, d_1 = -4.22035$ and $d_2 = 4067.87$ in the case of Fig. 2.  We see that
   the radial wave functions  rapidly decrease outside the well. Figures indicate some symmetry properties
   of the radial wave functions under replacement $m \to -(m+1)$.

In our opinion the examined  exactly solvable  model  with the realistic potential well of finite depth is physically
adequate in order to describe the behavior  of electron in a
semiconductor quantum dot with account of the Rashba spin-orbit
interaction. Further we intend to generalize our consideration by
including the magnetic field effects on the orbital motion of
electron in a quantum dot.

The author thanks K. Pankrashkin for pointing out the reference \cite{chap}.

\end{document}